\documentclass[twocolumn,showpacs,pre,floatfix,amsmath,amssymb,notitlepage]{revtex4-1}
\usepackage{epsfig}
\usepackage{subfigure}
\usepackage{amsmath}
\usepackage{color}
\usepackage{amssymb}
\usepackage{setspace}
\usepackage{graphicx}% Include figure files
\usepackage{dcolumn}% Align table columns on decimal point
\usepackage{bm}% bold math
\usepackage{times}
\usepackage{enumerate}
\usepackage{footnote}
\input epsf

\newcommand{\appropto}{\mathrel{\vcenter{
  \offinterlineskip\halign{\hfil$##$\cr
    \propto\cr\noalign{\kern2pt}\sim\cr\noalign{\kern-2pt}}}}}

\graphicspath{{figures/}}

\begin{document}

\author{Per Sebastian Skardal}
\email{persebastian.skardal@trincoll.edu} 
\affiliation{Department of Mathematics, Trinity College, Hartford, CT 06106, USA}

\title{Low Dimensional Dynamics of the Kuramoto Model with Rational Frequency Distributions}
%\date{\today}

\begin{abstract}
The Kuramoto model is a paradigmatic tool for studying the dynamics of collective behavior in large ensembles of coupled dynamical systems. Over the past decade a great deal of progress has been made in analytical descriptions of the macroscopic dynamics of the Kuramoto mode, facilitated by the discovery of Ott and Antonsen's dimensionality reduction method. However, the vast majority of these works relies on a critical assumption where the oscillators' natural frequencies are drawn from a Cauchy, or Lorentzian, distribution, which allows for a convenient closure of the evolution equations from the dimensionality reduction. In this paper we investigate the low dimensional dynamics that emerge from a broader family of natural frequency distributions, in particular a family of rational distribution functions. We show that, as the polynomials that characterize the frequency distribution increase in order, the low dimensional evolution equations become more complicated, but nonetheless the system dynamics remain simple, displaying a transition from incoherence to partial synchronization at a critical coupling strength. Using the low dimensional equations we analytically calculate the critical coupling strength corresponding to the onset of synchronization and investigate the scaling properties of the order parameter near the onset of synchronization. These results agree with calculations from Kuramoto's original self-consistency framework, but we emphasize that the low dimensional equations approach used here allows for a true stability analysis categorizing the bifurcations.
\end{abstract}


\maketitle

\section{Introduction}\label{sec:01}

Emergence of collective behavior in large ensembles of coupled dynamical units has long been an important area of research in the nonlinear dynamics and complexity communities~\cite{Pikovsky2003,Strogatz2004}. A particularly important model for exploring synchronization of coupled oscillators is the Kuramoto model~\cite{Kuramoto,Strogatz2000PhysD}, which serves as an analytically-tractable alternative to Winfree's earlier model~\cite{Winfree1967JTB}. Since its introduction the Kuramoto model and its variants have been the focus of a great deal of work aimed at uncovering analytical descriptions for the long-term macroscopic dynamics described by the so-called Kuramoto order parameter. Kuramoto first derived some initial results using a self-consistency analysis, shedding light on the critical coupling strength corresponding to the onset of synchronization as well as some calculations of the order parameter itself. However, this original analysis did not capture any stability properties of the system dynamics, making any formal bifurcation analysis unfeasible. Later, Strogatz and Mirollo made further progress in this direction, applying a Fokker-Planck approach that highlighted the first stability results for the system, specifically identifying the stability properties of the incoherent state~\cite{Strogatz1991JSP}.

Advances in the analytical description of the Kuramoto model's macroscopic dynamics have recently taken another significant leap with the work of Ott and Antonsen, who discovered a dimensionality reduction technique that allows for an explicit system of differential equations to be written down describing the dynamics of the order parameter itself~\cite{Ott2008Chaos,Ott2009Chaos}. This dimensionality reduction allows for a full stability and bifurcation analysis of the macroscopic system dynamics for the classical Kuramoto model as well as variants including effects such as time delays~\cite{Lee2009PRL,Skardal2018IJBC}, external forcing~\cite{Childs2008Chaos}, higher order coupling functions~\cite{Skardal2011PRE}, bimodal frequency distributions~\cite{Martens2009PRE}, community structure~\cite{Barreto2008PRE,Skardal2012PRE}, chimera states~\cite{Abrams2008PRL}, and complex network structure~\cite{Restrepo2014EPL,Skardal2015PRE}. In each of these cases, however, analytical progress is facilitated by the additional assumption that the distribution describing the oscillators' natural frequencies takes the form of a Cauchy, or Lorentzian, distribution. As we will see below, this specific choice is convenient precisely because of its simple rational form, allowing for the closure of the evolution equations under the so-called Ott-Antonsen ansatz~\cite{Ott2008Chaos}. 

In contrast to the large body of literature treating the Kuramoto model and its variants with Cauchy-distributed natural frequencies, few analytical results exist for more general frequency distributions. Moreover, other cases construct alternative frequency distributions by a superposition of multiple Cauchy distributions~\cite{Martens2009PRE,Omel2012PRL}, where similar techniques simplify the resulting evolution equations. In this work we generalize to the case of a generic family of rational frequency distributions. Specifically, we treat the classical Kuramoto model using the dimensionality reduction technique of Ott and Antonsen, but assume that the oscillators' natural frequencies are drawn from a more broad family of rational distribution functions (of which the Cauchy distribution is a special case). As the order of the polynomial characterizing the distribution increases, we find that the low-dimensional system describing the macroscopic system dynamics becomes more complicated. Specifically, when the frequency distribution decays like a polynomial of order $2n$, the macroscopic dynamics are governed by a system of $n$ complex-valued differential equations. Thus, the number of complex-valued ODEs that emerge from Ott and Antonsen's dimensionality reduction is precisely half the order of the distribution. Despite this growth in the complexity of the system of equations, we find that the dynamics are relatively simple, characterized by a single bifurcation from incoherence to partial synchronization at a critical coupling strength. We show that in addition to the low dimensional dynamics capturing fully the dynamics of the original system, the low dimensional equations can be used to analytically identify the critical coupling strength corresponding to the onset of synchronization. We also use the low dimensional equations to recover the scaling properties of the order parameter near the onset of synchronization. These results agree with those that one can obtain from Kuramoto's original self-consistency framework, but we emphasize that ere the low dimensional equations allows for a true bifurcation analysis that characterizes stability properties.

The remainder of this paper is organized as follows. In \ref{sec:02} we present the governing equations, briefly outline Ott and Antonsen's dimensionality reduction technique, and demonstrate the result for Cauchy-distributed frequencies. In \ref{sec:03} we introduce the more general family or rational frequency distributions we focus this work on and present some of its important properties, most notably the location of its poles in the complex plane and their residues. In \ref{sec:04} we apply the dimensionality reduction technique to family of frequency distributions introduced in \ref{sec:03}. First we consider the generic family of rational distributions, then consider the specific cases of quartic, sextic, and octic distributions. In each case we show that the low dimensional dynamics capture the original system dynamics and perform a linear stability analysis of the incoherent state, used to identify the onset of synchronization. In \ref{sec:05} we use our results to study the scaling properties of the system near the onset of sycnhronization. Finally, in \ref{sec:06} we conclude with a discussion of our results.

\section{Model Description, Dimensionality Reduction, and the Cauchy Distribution}\label{sec:02}

In this work we will focus our attention on the macroscopic dynamics of the classical Kuramoto model. Although several variants that incorporate addition features have been considered, e.g., time delayed interactions, network structure, etc., we study Kuramoto's original formulation consisting of a system of $N$ oscillators whose states are described, respectively, by the angles $\theta_n\in[0,2\pi)$ and evolve according to
\begin{align}
\dot{\theta}_n=\omega_n+\frac{K}{N}\sum_{m=1}^N\sin(\theta_m-\theta_n),\label{eq:01}
\end{align}
where $K\ge0$ is the global coupling strength and $\omega_n$ is the natural frequency of oscillator $n$, assumed to be drawn independently and identically from the distribution function $g(\omega)$. The dynamics of each oscillator in the Kuramoto model depend on two system parameters: the coupling strength $K$, which is shared by all oscillators in the system, and its natural frequency $\omega_n$, which is specific to oscillator $n$. Throughout the whole system the natural frequencies are described by the distribution function $g(\omega)$, and therefore the collective dynamics of the Kuramoto model, i.e., the degree to which the system displays synchronization or incoherence, depends on the interplay between $K$ and $g(\omega)$. Broadly speaking, if $K$ is large enough with respect to the spread of $g(\omega)$, the system will typically synchronize. 

Next, to measure the degree of synchronization in the system we use Kuramoto's order parameter given by the complex number $z$ defined as follows:
\begin{align}
z =re^{i\psi}=\frac{1}{N}\sum_{m=1}^Ne^{i\theta_m}.\label{eq:02}
\end{align}
The Kuramoto order parameter given in Eq.~(\ref{eq:02}) represents the centroid within the complex unit disc of all oscillators placed at their respective angle on the complex unit circle. The degree of synchronization in the system is most conveniently expressed using the polar decomposition $z=re^{i\psi}$, where $r$ is the magnitude of $z$ and $\psi$ is the collective phase. Note that in the case where oscillators are spread roughly uniformly along the unit circle we have that $r\approx0$, while in the case where a large number of oscillator are tightly concentrated around a common phase we have that $r\approx1$. Thus, for large system, i.e., $N\gg1$, we interpret $r\approx0$ as an incoherent state and $r>0$ as a partially synchronized state. Moreover, Eq.~(\ref{eq:02}) allows us to rewrite the dynamics in Eq.~(\ref{eq:01}) in a more simple form as
\begin{align}
\dot{\theta}_n&=\omega_n+Kr\sin(\psi-\theta_n)\nonumber\\
&=\omega_n+\frac{K}{2i}\left(ze^{i\theta_n}-z^*e^{-i\theta_n}\right),\label{eq:03}
\end{align}
from which we may interpret each oscillator as reacting not necessarily to each other oscillator individually, but collectively through a single mean field described by the order parameter.

We now briefly outline the dimensionality reduction technique discovered by Ott and Antonsen. First we consider the continuum limit $N\to\infty$, whereby the overall state of the system can be described by a distribution of oscillators, denoted $f(\omega,\theta,t)$, where $f(\omega,\theta,t)d\omega d\theta$ represents the fraction of oscillators with natural frequency between $\omega$ and $\omega+d\omega$ and phase between $\theta$ and $\theta+d\theta$ at time $t$. First, we note that by the conservation of oscillators, $f$ must satisfy the following continuity equation
\begin{align}
\frac{\partial}{\partial t} f + \frac{\partial}{\partial \theta}\left(\dot{\theta}f\right)=0,\label{eq:04}
\end{align}
which contains no $\partial/\partial \omega$ term since natural frequencies are fixed. Moreover, the Fourier series of $f$ must be of the form
\begin{align}
f(\omega,\theta,t)=\frac{g(\omega)}{2\pi}\left[1+\sum_{n=1}^\infty \hat{f}_n(\omega,t)e^{in\theta}+c.c.\right],\label{eq:05}
\end{align}
where $\hat{f}_n(\omega,t)$ is the $n^{\text{th}}$ coefficient of the Fourier series and $c.c.$ denotes the complex conjugate of the previous term. Note that in this form, determining the state of the system via the distribution $f$ is equivalent to determining the set of Fourier coefficients $\hat{f}_n$. The main discovery by Ott and Antonsen begins with an ansatz that collapses all of the Fourier coefficients onto one condition. Specifically, by proposing that the Fourier coefficients decay geometrically, i.e., $\hat{f}_n(\omega,t)=\alpha^n(\omega,t)$ for some function $\alpha(\omega,t)$, and inserting Eq.~(\ref{eq:05}) into Eq.~(\ref{eq:04}), the evolution of $\alpha$, and thereby the distribution $f$, is determined by the single differential equation
\begin{align}
\frac{\partial}{\partial t}\alpha + i\omega\alpha + \frac{K}{2}\left(z\alpha^2-z^*\right)=0,\label{eq:06}
\end{align}
where $*$ represents the complex conjugate. To close the dynamics of the system we use that in the continuum limit we have that $z(t)=\iint e^{i\theta}f(\omega,\theta,t)d\theta d\omega$. Inserting the Fourier series along with the ansatz considered above, we have that
\begin{align}
z^*=\int_{-\infty}^\infty\alpha(\omega,t)g(\omega)d\omega.\label{eq:07}
\end{align}

Together, Eqs.~(\ref{eq:06}) and (\ref{eq:07}) close the macroscopic dynamics of the system, where Eq.~(\ref{eq:06}) governs the evolution of $\alpha$ in response to $z$ and the coupling strength $K$, while Eq.~(\ref{eq:07}) determines $z$ as a function of $\alpha$ and the frequency distribution $g$. In the vast majority of works that have utilized the dimensionality reduction of Ott and Antonsen, an addition assumption is made on the frequency distribution $g(\omega)$ to simplify the dynamics further. Specifically, it is often assumed that $g(\omega)$ is given by a Cauchy, or Lorentzian, distribution of the form
\begin{align}
g(\omega)=\frac{\Delta}{\pi\left[(\omega-\Omega)^2+\Delta^2\right]},\label{eq:08}
\end{align}
with mean $\Omega$ and spread $\Delta>0$. (We note that by entering the rotating frame $\theta\mapsto\theta+\Omega t$ in Eq.~(\ref{eq:01}) we may set the mean $\Omega=0$ without loss of generality. For the remainder of this paper we will therefore assume that the mean of any distribution we consider has mean $\Omega=0$.) Note that Eq.~(\ref{eq:08}) has two simple poles located at $\omega=\pm i\Delta$. By assuming that the function $\alpha(\omega,t)$ satisfies (i) $|\alpha(\omega,t)|\le1$, (ii) $\alpha(\omega,0)$ is analytically continuable into the lower-half complex $\omega$ plane, and (iii) $\alpha(\omega,t)\to0$ as $\text{Im}(\omega)\to-\infty$, Eq.~(\ref{eq:07}) can be evaluated analytically by closing the integration along the semicircle of infinite radius in the lower-half plane and using Cauchy's residue theorem~\cite{Ablowitz2003} with the pole at $\omega=-i\Delta$. In particular, denoting $\text{Res}(g;-i\Delta)$ as the residue of $g(\omega)$ at $\omega=-i\Delta$, we have that
\begin{align}
z^*=(-2\pi i) \alpha(-i\Delta,t) \text{Res}(g;-i\Delta)=\alpha(-i\Delta,t).\label{eq:09}
\end{align}
This allows for the closure of the macroscopic dynamics within a single complex differential equation by evaluating Eq.~(\ref{eq:06}) at the pole $\omega=-i\Delta$ and taking a complex conjugate, yielding
\begin{align}
\dot{z}=-\Delta z+\frac{K}{2}\left(z-z^*z^2\right).\label{eq:10}
\end{align}
Using the polar decomposition $z=re^{i\psi}$, Eq.~(\ref{eq:10}) can be transformed to the system of two real-valued equations,
\begin{align}
\dot{r}&=-\Delta r + \frac{K}{2}\left(r-r^3\right),\label{eq:11}\\
\dot{\psi} &=0.\label{eq:12}
\end{align}
Searching for a stable fixed-point of Eq.~(\ref{eq:11}) yields the long-term macroscopic state given by
\begin{align}
r = \left\{\begin{array}{rl}0 &\text{if }K\le2\Delta,\\ \sqrt{1-\frac{2\Delta}{K}} &\text{if }K>2\Delta,\end{array}\right.\label{eq:13}
\end{align}
where a pitchfork bifurcation occurs at the critical coupling strength $K_c=2\Delta$ indicating a transition from incoherence, $r=0$, to partial synchronization, $r>0$, i.e., the onset of synchronization. 

\begin{figure}[t]
\centering
\epsfig{file =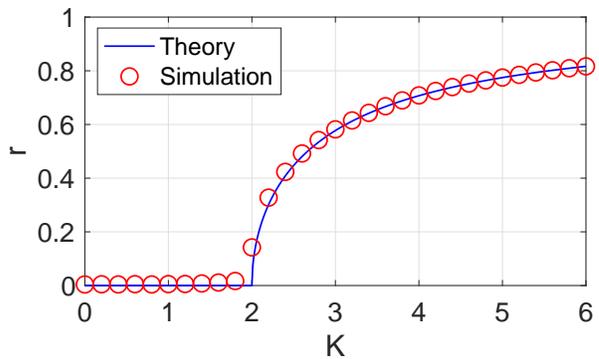, clip =,width=1.0\linewidth }
\caption{{\bf Macroscopic steady-state: Cauchy-distributed natural frequencies}. Steady-state of the macroscopic dynamics measured by the amplitude of the order parameter, $r=|z|$, for Cauchy-distributed natural frequencies with $\Delta=1$. Results from a direct simulation of a system of $N=50000$ oscillators is illustrated with red circles, while the analytical prediction given in Eq.~(\ref{eq:13}) is illustrated with a blue curve.}\label{fig:01}
\end{figure}

In Fig.~\ref{fig:01} we illustrate the ability for the low dimensional analysis to capture the original system dynamics, plotting the analytical prediction given in Eq.~(\ref{eq:13}) in blue, compared to results from direct simulation of a system with $N=50000$ oscillators using red circles. (Here we use $\Delta=1$.) We note that the agreement between theory and direct simulation is excellent. We turn now to investigate the macroscopic dynamics that emerge from different choices of the distribution function $g(\omega)$. 

\section{Rational Frequency Distributions}\label{sec:03}

Our aim in this work is to generalize the typical choice of using Cauchy-distributed natural frequencies to a broader family of distributions. Note that the choice of a Cauchy distribution, i.e., Eq.~(\ref{eq:08}), simplifies the dynamics of Eqs.~(\ref{eq:06}) and (\ref{eq:07}) to Eq.~(\ref{eq:10}) precisely because it is a rational polynomial with two simple poles, one of which lies in the bottom-half complex plane. On the other hand, other choices of $g(\omega)$ may still allow for analytical evaluation of Eq.~(\ref{eq:07}) provided that it is appropriately rational, i.e., has simple poles away from the real line. In fact, in their original work Ott and Antonsen speculated at this possibility~\cite{Ott2008Chaos}. With this in mind we consider a broader family of frequency distributions for $g(\omega)$, specifically the following family of distributions parameterized by the integer $n\ge1$:
\begin{align}
g_n(\omega)=n\sin(\pi/2n)\frac{\Delta^{2n-1}}{\pi\left(\omega^{2n}+\Delta^{2n}\right)},\label{eq:14}
\end{align}
where $\Delta>0$ is the spread parameter. (We note that a similar distribution was considered in Ref.~\cite{Lafuerza2010PRL}.)

\begin{figure}[t]
\centering
\epsfig{file =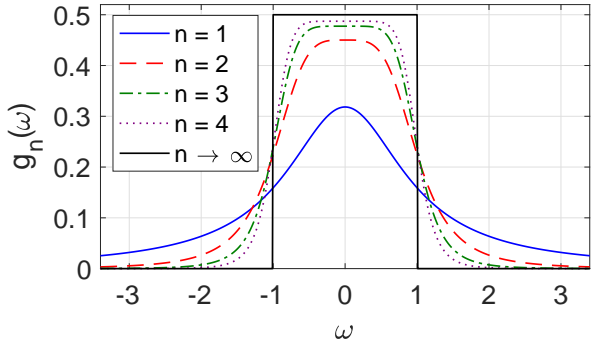, clip =,width=1.0\linewidth }
\caption{{\bf Rational natural frequency distriutions}. An illustration of the rational natural frequency distributions $g_n(\omega)$ defined in Eq.~(\ref{eq:14}) for $n=1$, $2$, $3$, $4$, and the limit $n\to\infty$ (black box), i.e., the quadratic, quartic, sextic, octic, and uniform distributions, respectively.}\label{fig:02}
\end{figure}

First, note that the choice $n=1$ recovers precisely the Cauchy distribution. Thus Eq.~(\ref{eq:14}) represents a natural generalization of the Cauchy distribution. Moreover, we adopt the nomenclature of the order of the polynomial in the denominator, so that we refer to the $n=1$ case as the quadratic distribution, $n=2$ as the quartic distribution, etc. Second, Eq.~(\ref{eq:14}) is a convenient choice because it can be used to represent the uniform distribution, which has been used in several studies involving the Kuramoto model, e.g., Ref.~\cite{Pazo2005PRE}. In particular, taking the limit $n\to\infty$ and using the series expansion for $\sin(\pi/2n)$ yields
\begin{align}
\lim_{n\to\infty}g_n(\omega)&=\lim_{n\to\infty}n\sin(\pi/2n)\frac{\Delta^{2n-1}}{\pi\left(\omega^{2n}+\Delta^{2n}\right)}\nonumber\\
&= \lim_{n\to\infty}n\left(\sum_{m = 0}^\infty\frac{(\pi/2n)^{2m+1}}{(2m+1)!}\right)\nonumber\\
&\hskip16ex\times\frac{\Delta^{-1}}{\pi\left(\left(\omega/\Delta\right)^{2n}+1\right)}\nonumber\\
&=\lim_{n\to\infty}\frac{\pi}{2}\frac{\Delta^{-1}}{\pi\left(\left(\omega/\Delta\right)^{2n}+1\right)}\nonumber\\
&=\left\{\begin{array}{rl}0 &\text{if }\omega>\Delta,\\ 1/(4\Delta)&\text{if }\omega=\Delta, \\ 1/(2\Delta)&\text{if }\omega<\Delta.\end{array}\right.\label{eq:15}
\end{align}
Thus, the case of uniformly-distributed frequencies can be treated by considering the $n\to\infty$ limit of the rational distribution given in Eq.~(\ref{eq:14}). In Fig.~\ref{fig:02} we illustrate the shape of $g_n(\omega)$ as $n$ is varied, plotting the quadratic, quartic, sextic, and octic cases, i.e., $n=1$, $2$, $3$, and $4$ in solid blue, dashed red, dot-dashed green, and  and dotted purple, with the $n\to\infty$ limit resulting in the uniform distribution illustrated as the black box. 

Our interest lies in the effect that different choices of the distribution function $g(\omega)=g_n(\omega)$ for $n=1,2,\dots$ have on the resulting macroscopic dynamics that emerge from the dimensionality reduction of Ott and Antonsen. This requires the unification of Eq.~(\ref{eq:06}) and Eq.~(\ref{eq:07}) via analytically integrating Eq.~(\ref{eq:07}) using Cauchy's residue theorem. Thus, to close the low dimensional dynamics we require the poles of each $g_n(\omega)$ and their respective residues.

\begin{figure*}[t]
\centering
\epsfig{file =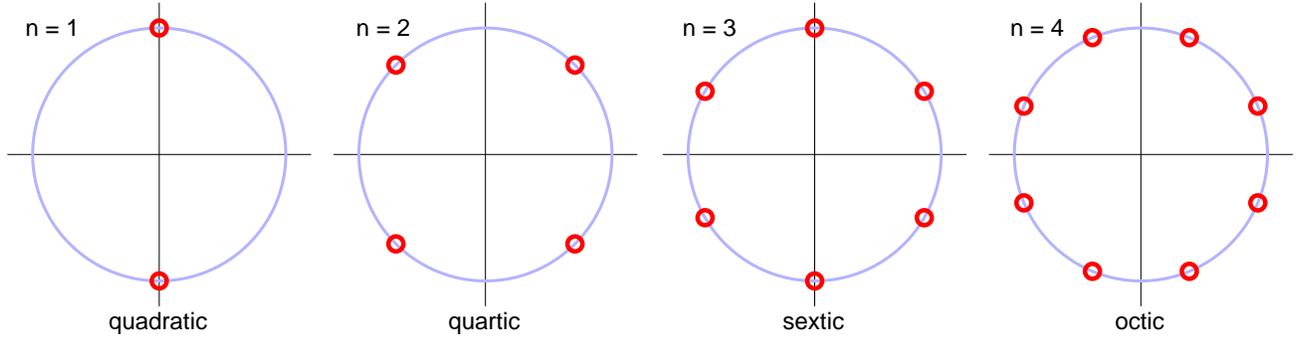, clip =,width=1.0\linewidth }
\caption{{\bf Poles of Rational functions}. Illustration of the poles $\widehat{\omega}$ of the rational function $g_n(\omega)=n\sin(\pi/2n)\Delta^{2n-1}/[\pi(\omega^{2n}+\Delta^{2n})]$ for $n=1$, $2$, $3$, and $4$, i.e., the quadratic, quartic, sextic, and octic cases. Poles are given by Eq.~(\ref{eq:21}) and (\ref{eq:20}), all lying on the circle $|\widehat{\omega}|=\Delta$.}\label{fig:03}
\end{figure*}

First, to find the location of the poles, we note that from Eq.~(\ref{eq:14}) we see that the poles, denoted $\hat{\omega}$, occur when the denominator vanishes, i.e., satisfy
\begin{align}
\widehat{\omega}^{2n}+\Delta^{2n}=0.\label{eq:16}
\end{align}
Considering complex numbers in polar form $\widehat{\omega}=\rho e^{i\phi}$ (note that $\widehat{\omega}=0$ is not a pole) we have that
\begin{align}
\rho^{2n}e^{2ni\phi}=-\Delta^{2n},\label{eq:17}
\end{align}
and after taking an absolute value we find that $\rho=\Delta$ and 
\begin{align}
e^{2ni\phi}=-1,\label{eq:18}
\end{align}
or equivalently,
\begin{align}
2n\phi=\pm\left(\pi+2k\pi\right),\label{eq:19}
\end{align}
where $k\in\mathbb{Z}$. Dividing through by $2n$ yields
\begin{align}
\phi=\pm\left(\frac{\pi}{2n}+\frac{k\pi}{n}\right)=\pm\frac{(2k+1)\pi}{2n}.\label{eq:20}
\end{align}
Thus, the poles of $g_n(\omega)$ are located at 
\begin{align}
\widehat{\omega}=\Delta e^{i\phi},\label{eq:21}
\end{align}
where $\phi$ is given in Eq.~(\ref{eq:20}) for $k=0,1,\dots,n-1$. Note that this reveals some very specific symmetries. First, poles are located along equally-spaced intervals on the circle of radius $\Delta$. Moreover, there are precisely $2n$ poles, occurring in complex conjugate pairs: $n$ in the upper-half plane and $n$ in the lower-half plane. Since the left hand side of Eq.~(\ref{eq:16}) is a polynomial of order $2n$ and $2n$ distinct poles exist this implies that each pole is simple. In Fig.~\ref{fig:03} we illustrate the location of the poles of $g_n(\omega)$, plotting them in the complex plane for the cases $n=1$, $2$, $3$, and $4$, i.e., the quadratic, quartic, sextic, and octic cases. We denote the circle of radius $\Delta$, i.e., $|\omega|=\Delta$, in blue, making clear that all poles, illustrated by red circles, are evenly spaced along this circle with a complex conjugate symmetry.

In addition to the poles themselves, we also require the respective Residues of $g_n(\omega)$ at each of the poles $\widehat{\omega}$. Recall that, given the complex function $f(z)$ with a pole at $\widehat{z}$, the residue of $f(z)$ at $z=\widehat{z}$, denoted $\text{Res}(f;\widehat{z})$ is given by the coefficient of the $1/(z-\widehat{z})$ term in the Laurent series of $f(z)$ about $z=\widehat{z}$~\cite{Ablowitz2003}. As we have seen above, the poles of each $g_n$ are all simple, so we may calculate the residues via
\begin{align}
\text{Res}(g_n;\omega)=\lim_{z\to\widehat{z}}g_n(\omega)\cdot(\omega-\widehat{\omega}),\label{eq:22}
\end{align}
or equivalently
\begin{align}
\text{Res}(g_n;\omega)=\varphi_{n,\widehat{\omega}}(\omega)\big|_{\omega=\widehat{\omega}},\label{eq:23}
\end{align}
where $\varphi_{n,\widehat{\omega}}(\omega)$ is the result of applying polynomial division to $g_n(\omega)\cdot(\omega-\widehat{\omega})$. Here we calculate the full set of residues $\text{Res}(g_n;\widehat{\omega})$ for the quadratic, quartic, sextic, and octic cases, i.e., $n=1$, $2$, $3$, and $4$, and organize them in Table~\ref{table:Residue} in Appendix~\ref{app:A}. In the following section we make use of them in order to write down the low dimensional dynamics for these choices of natural frequency distributions.

\section{Low Dimensional Dynamics using Rational Frequency Distributions}\label{sec:04}

We now turn our attention to the dynamics of the Kuramoto model for various instances of rational frequency distributions taken from the family defined in Eq.~(\ref{eq:14}). In particular, we consider the dynamics given in Eq.~(\ref{eq:01}) in the continuum limit, $N\to\infty$, whose dynamics after applying the dimensionality reduction of Ott and Antonsen are determined by Eq.~(\ref{eq:06}) and Eq.~(\ref{eq:07}). We then seek to eliminate the integral in Eq.~(\ref{eq:07}) by analytical evaluation. However, because we now consider natural frequency distributions that are not Cauchy this becomes more complicated. We first consider the general case of arbitrary integers $n\ge1$, then consider some specific choices as examples. 

To evaluate Eq.~(\ref{eq:07}) analytically for general $g_n(\omega)$, we proceed as we before, by considering the closed path $C$ in the complex plane consisting of the real line, denoted $C_1$, oriented from negative real part to positive real part, together with the semicircle of infinite radius in the lower-half complex plane, denoted $C_2$, oriented in the clockwise (i.e., negative) direction. Noting that integration along $C_1$ corresponds to integration over the real number, we then have that
\begin{align}
\int_{-\infty}^\infty\alpha(\omega,t)g(\omega)d\omega&=\oint_{C}\alpha(\omega,t)g(\omega)d\omega\nonumber\\
&\hskip2ex-\int_{C_2}\alpha(\omega,t)g(\omega)d\omega.\label{eq:24}
\end{align}
Moreover, recall that in Sec.~\ref{sec:02} three important assumptions were made regarding $\alpha(\omega,t)$, namely (i) $|\alpha(\omega,t)|\le1$, (ii) $\alpha(\omega,0)$ is analytically continuable into the lower-half complex $\omega$ plane, and (iii) $\alpha(\omega,t)\to0$ as $\text{Im}(\omega)\to-\infty$. Under these conditions the latter integral on the right hand side of Eq.~(\ref{eq:24}) vanishes, resulting in
\begin{align}
z^*=\oint_{C}\alpha(\omega,t)g(\omega)d\omega,\label{eq:25}
\end{align}
to which we may apply Cauchy's residue theorem. To do so, we make two important observations. First, the closed loop $C$ is negatively-oriented. Second, because $C$ encloses the lower-half complex plane, it contains precisely the $n$ poles with negative imaginary part associated with the respective distribution $g_n(\omega)$. The closed path integral can then be evaluated, resulting in
\begin{align}
z^*=-2\pi i\sum_{k=0}^{n-1}\alpha(\widehat{\omega}_k,t)\text{Res}(g_n;\widehat{\omega}_k),\label{eq:26}
\end{align}
where $\widehat{\omega}_k$ is one of the poles given in Table~\ref{table:Residue} in Appendix~\ref{app:A}, given by
\begin{align}
\widehat{\omega}_k=\Delta\text{exp}\left(\frac{-(2k+1)\pi i}{2n}\right),\label{eq:27}
\end{align}
for $k=0,\dots,n-1$. Next, it is convenient to define a set of pseudo order parameters $z_k$, as
\begin{align}
z_{k}^*=-2\pi i\text{Res}(g_n;\widehat{\omega}_k)\alpha(\widehat{\omega}_k,t),\label{eq:28}
\end{align}
allowing us to write
\begin{align}
z^* = \sum_{k=0}^{n-1}z_k^*.\label{eq:29}
\end{align}
Because $z$ is a combination of the pseudo order parameters $z_k$, i.e., the function $\alpha(\omega,t)$ evaluated at different poles $\widehat{\omega}_k$, a single differential equation for $z$ cannot be obtained directly. However, a distinct differential equation can be obtained for each different $z_k$ by evaluating Eq.~(\ref{eq:06}) at the respective pole. In particular, evaluating at $\omega=\widehat{\omega}_k$ and taking a complex conjugate yields
\begin{widetext}
\begin{align}
\dot{z}_k= i\widehat{\omega}_k^* z_k + \frac{K}{2}\left[2\pi i\text{Res}^*(g_n;\widehat{\omega}_k)\left(\sum_{l=0}^{n-1}z_l\right)-\left(\sum_{l=0}^{n-1}z_l^*\right)\frac{z_k^{2}}{2\pi i\text{Res}^*(g_n;\widehat{\omega}_k)}\right].\label{eq:30}
\end{align}
\end{widetext}
Equation~(\ref{eq:30}) can be obtained by evaluating for each pole $\widehat{\omega}_k$, thereby giving one differential equation for each pseudo order parameter $z_k$ for $k=0,\dots,n-1$. This is similar to the cases of multi-modal frequency distributions, e.g., \cite{Martens2009PRE}, however in this case the underlying frequency distribution is not multimodal and the number of differential equations produced depends on the order of the frequency distribution. Next we will investigate the structure and dynamics of the resulting systems for some particular choices of frequency distributions, specifically the quartic, sextic, and octic distributions, noting that the quadratic distribution is precisely the Cauchy distribution considered as an introductory example in Sec.~\ref{sec:02}.

\subsection{Quartic Distribution}\label{subsec:0401}

We begin with the case of the quartic distribution, i.e., $g(\omega)=g_n(\omega)$ for $n=2$. From Table~\ref{table:Residue} there are two poles in the lower-half complex plane given by
\begin{align}
\widehat{\omega}_0=\frac{1}{\sqrt{2}}\Delta-\frac{1}{\sqrt{2}}\Delta i,\hskip2ex\text{and}\hskip2ex\widehat{\omega}_1=-\frac{1}{\sqrt{2}}\Delta-\frac{1}{\sqrt{2}}\Delta i,\label{eq:31}
\end{align}
whose residues are, respectively
\begin{align}
\text{Res}(g_2;\widehat{\omega}_0)=\frac{-1-i}{4\pi i},\hskip2ex\text{and}\hskip2ex\text{Res}(g_2;\widehat{\omega}_1)=\frac{-1+i}{4\pi i}.\label{eq:32}
\end{align}
Inserting these into Eq.~(\ref{eq:30}) for both poles yields the following system of two complex-valued differential equations:
\begin{align}
\dot{z}_0 &=-\frac{\left(1-i\right)\Delta}{\sqrt{2}}z_0+\frac{K}{4}\left[(1-i)\left(z_0+z_1\right)-\frac{4\left(z_0^*+z_1^*\right)z_0^2}{1-i}\right], \label{eq:33}\\
\dot{z}_1 &=-\frac{\left(1+i\right)\Delta}{\sqrt{2}}z_1+\frac{K}{4}\left[(1+i)\left(z_0+z_1\right)-\frac{4\left(z_0^*+z_1^*\right)z_1^2}{1+i}\right], \label{eq:34}
\end{align}
with the Kuramoto order parameter defined by $z=z_0+z_1$.

\begin{figure}[t]
\centering
\epsfig{file =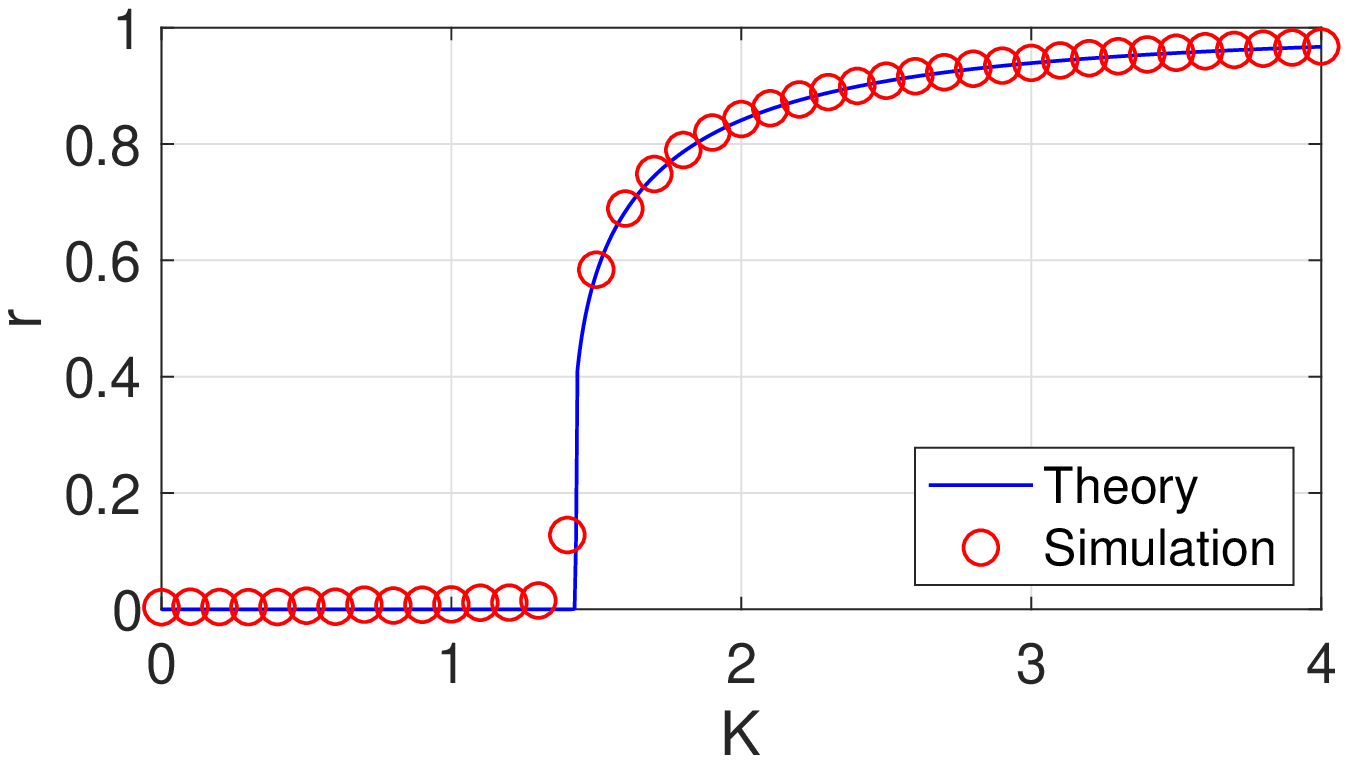, clip =,width=1.0\linewidth }
\caption{{\bf Macroscopic steady-state: Quartic distribution}. Steady-state of the macroscopic dynamics measured by the amplitude of the order parameter, $r=|z|$, for the quartic distribution, i.e., $n=2$, with $\Delta=1$. Results from a direct simulation of a system of $N=50000$ oscillators is illustrated with red circles, while the analytical prediction obtained by simulating the low dimensional dynamics in Eqs.~(\ref{eq:33}) and (\ref{eq:34}), where $z=z_0+z_1$, is illustrated with a blue curve.}\label{fig:04}
\end{figure}

We begin by illustrating the utility of capturing the macroscopic dynamics of large systems using the low dimensional dynamics of Eqs.~(\ref{eq:33}) and (\ref{eq:34}). In Fig.~\ref{fig:04} we plot the degree of synchronization $r=|z|$ vs the coupling strength $K$ obtain via two methods. First, we use simulations of the low dimensional system in Eqs.~(\ref{eq:33}) and (\ref{eq:34}) with $\Delta=1$, denoting results using a blue curve. Second, we use direct simulations of the system in Eq.~(\ref{eq:01}) with $N=50000$ oscillators whose frequencies are drawn from the quartic distribution, i.e., $n=2$, with $\Delta=1$, denoting results with red circles. We note an excellent agreement between the theoretical and directly simulated results.

Next we turn our attention to the onset of synchronization, characterizing the bifurcation between the loss of stability of the incoherent state, $r=0$, and the appearance of the partially synchronized state, $r>0$. In the case of Cauchy-distributed natural frequencies, we see from Eq.~(\ref{eq:13}) that this comes in the form of a pitchfork bifurcation at $K_c=2\Delta$. However, we see from Fig.~\ref{fig:04} that for a quartic distribution this critical value must be smaller, i.e., the onset of synchronization occurs earlier. To find this, we conduct a linear stability analysis of the incoherent state $z=0$. In particular, since the low dimensional dynamics in Eqs.~(\ref{eq:33}) and (\ref{eq:34}) are given in terms of $z_0$ and $z_1$, and $z=z_0+z_1$, we conduct a linear stability analysis about the $z_0=0$, $z_1=0$ state. (We note that it is possible to have $z_0,z_1\ne0$ and $z=0$, but we find through numerical simulations of the low dimensional equations that this does not occur, and the loss of stability of $(z_0,z_1)=(0,0)$ does in fact correspond to the loss of stability of $z=0$.)

The linear stability analysis of the incoherent state is most simply done by expressing Eqs.~(\ref{eq:33}) and (\ref{eq:34}) in real and imaginary parts, i.e., expressing the differential equations for $x_0$, $y_0$, $x_1$, and $y_1$, where $z_0=x_0+iy_0$ and $z_1=x_1+iy_1$, and evaluating the new system Jacobian at $(x_0,y_0,x_1,y_1)=(0,0,0,0)$. This results in the following $4\times4$ Jacobian matrix
\begin{widetext}
\begin{align}
DF=\frac{1}{4}\begin{bmatrix}K-2\sqrt{2}\Delta & K-2\sqrt{2}\Delta & K & K\\ -K+2\sqrt{2}\Delta & K-2\sqrt{2}\Delta & -K & K\\ K & -K & K-2\sqrt{2}\Delta & -K+2\sqrt{2}\Delta\\ K & K & K-2\sqrt{2}\Delta & K-2\sqrt{2}\Delta\end{bmatrix},\label{eq:35}
\end{align}
whose eigenvalues are given by
\begin{align}
\lambda_{1,2} = \frac{1}{4}\left(K-2\sqrt{2}\Delta-\sqrt{K^2+4\sqrt{2}K\Delta-8\Delta^2}\right),\hskip2ex\lambda_{3,4} = \frac{1}{4}\left(K-2\sqrt{2}\Delta+\sqrt{K^2+4\sqrt{2}K\Delta-8\Delta^2}\right),\label{eq:36}
\end{align}
\end{widetext}
where the degeneracy of repeated eigenvalues indicates the collective rotational invariance $\theta\mapsto\theta+\delta\theta$ in the original system Eq.~(\ref{eq:01}). To find the onset of synchronization we search for the coupling strength where the $z_0,z_1=0$ state first becomes unstable, i.e., when the largest real part of the set of eigenvalues crosses $\text{Re}(\lambda)=0$ into the right-half complex plane. In principle, this can occur with either complex eigenvalues, i.e., $\text{Im}(\lambda)\ne0$, corresponding to a Hopf bifurcation, or real eigenvalues, i.e., $\text{Im}(\lambda)=0$, corresponding to a pitchfork bifurcation. Here we find that the onset of synchronization occurs via the latter mechanism, whereby setting $\lambda_{3,4}=0$ yields the onset of synchronization at $K_c=\sqrt{2}\Delta$, which is in agreement with the results plotted in Fig.~\ref{fig:04}. 

Before continuing on to a new case, we note that this is in agreement with Kuramoto's original self-consistency analysis. Specifically, for a unimodal natural frequency distribution $g(\omega)$ that is symmetric about the mean $\omega=\Omega$, the onset of synchronization is given by $K_c=2/(\pi g(\Omega))$~\cite{Kuramoto}.

\subsection{Sextic Distribution}\label{subsec:0402}

Next we consider the case of the sextic distribution, i.e., $g(\omega)=g_n(\omega)$ for $n=3$. From Table~\ref{table:Residue} there are three poles in the lower-half complex plane given by
\begin{widetext}
\begin{align}
\widehat{\omega}_0=\frac{\sqrt{3}}{2}\Delta-\frac{1}{2}\Delta i,\hskip2ex\widehat{\omega}_1=-\Delta i,\hskip2ex\text{and}\hskip2ex \widehat{\omega}_2=-\frac{\sqrt{3}}{2}\Delta-\frac{1}{2}\Delta i\label{eq:37}
\end{align}
whose residues are, respectively
\begin{align}
\text{Res}(g_3;\widehat{\omega}_0)=\frac{-1-\sqrt{3}i}{8\pi i},\hskip2ex\text{Res}(g_3;\widehat{\omega}_1)=\frac{-1}{4\pi i},\hskip2ex\text{and}\hskip2ex\text{Res}(g_3;\widehat{\omega}_2)\frac{-1+\sqrt{3}i}{8\pi i}.\label{eq:38}
\end{align}
Inserting these into Eq.~(\ref{eq:30}) for both poles yields the following system of three complex-valued differential equations:
\begin{align}
\dot{z}_0 &=-\frac{\left(1-\sqrt{3}i\right)\Delta}{2}z_0+\frac{K}{8}\left[(1-\sqrt{3}i)\left(z_0+z_1+z_2\right)-\frac{16\left(z_0^*+z_1^*+z_2^*\right)z_0^2}{1-\sqrt{3}i}\right], \label{eq:39}\\
\dot{z}_1 &=-\Delta z_1+\frac{K}{4}\left[\left(z_0+z_1+z_2\right)-4\left(z_0^*+z_1^*+z_2^*\right)z_1^2\right], \label{eq:40}\\
\dot{z}_2 &=-\frac{\left(1+\sqrt{3}i\right)\Delta}{2}z_2+\frac{K}{8}\left[(1+\sqrt{3}i)\left(z_0+z_1+z_2\right)-\frac{16\left(z_0^*+z_1^*+z_2^*\right)z_2^2}{1+\sqrt{3}i}\right], \label{eq:41}
\end{align}
\end{widetext}
with the Kuramoto order parameter defined by $z=z_0+z_1+z_2$.

\begin{figure}[htbp]
\centering
\epsfig{file =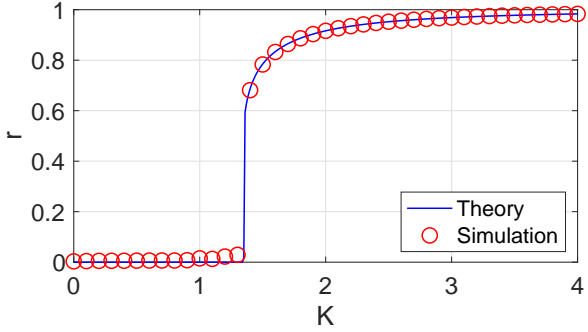, clip =,width=1.0\linewidth }
\caption{{\bf Macroscopic steady-state: Sextic distribution}. Steady-state of the macroscopic dynamics measured by the amplitude of the order parameter, $r=|z|$, for the sextic distribution, i.e., $n=3$, with $\Delta=1$. Results from a direct simulation of a system of $N=50000$ oscillators is illustrated with red circles, while the analytical prediction obtained by simulating the low dimensional dynamics in Eqs.~(\ref{eq:39}), (\ref{eq:40}), and (\ref{eq:41}), where $z=z_0+z_1+z_2$, is illustrated with a blue curve.}\label{fig:05}
\end{figure}

We again illustrate the utility of capturing the macroscopic dynamics of large systems using the low dimensional dynamics of Eqs.~(\ref{eq:39}), (\ref{eq:40}), and (\ref{eq:41}). In Fig.~\ref{fig:05} we plot the degree of synchronization $r=|z|$ vs the coupling strength $K$ obtained first using simulations of the low dimensional system in Eq.~(\ref{eq:39}), (\ref{eq:40}), and (\ref{eq:41}) with $\Delta=1$, denoting results using a blue curve, and second using direct simulations of the system in Eq.~(\ref{eq:01}) with $N=50000$ oscillators whose frequencies are drawn from the sextic distribution, i.e., $n=3$, with $\Delta=1$, denoting results with red circles. Again, we note an excellent agreement between the theoretical and directly simulated results.

Next, we illustrate that a linear stability analysis may in this case also reveal the onset of synchronization. Similar to the quartic case, we investigate the stability of the $(z_0,z_1,z_2)=(0,0,0)$ state by calculating the Jacobian matrix of the dynamics after separating into real and imaginary parts, $z_0=x_0+iy_0$, $z_1=x_1+iy_1$, and $z_2=x_2+iy_2$. The resulting $6\times 6$ Jacobian matrix for $(x_0,y_0,x_1,y_1,x_2,y_2)=(0,0,0,0,0,0)$ is given by
\begin{align}
DF=\frac{1}{8}\begin{bmatrix}
a & \sqrt{3}a & K & \sqrt{3}K & K & \sqrt{3}K\\
-\sqrt{3}a & a & -\sqrt{3}K & K & -\sqrt{3}K & K\\
2K & 0 & 2a & 0 & 2K & 0\\
0 & 2K & 0 & 2a & 0 & 2K\\
K & -\sqrt{3}K & K & -\sqrt{3}K & a & -\sqrt{3}a\\
\sqrt{3}K & K & \sqrt{3}K & K & \sqrt{3}a & a
 \end{bmatrix},\label{eq:42}
\end{align}
where $a=K-4\Delta$. The eigenvalues of $DF$ in Eq.~(\ref{eq:42}) are difficult to write down in closed form, but are given by the roots of the polynomial $\rho(\lambda)$ given by
\begin{align}
\rho(\lambda)&=4\lambda^3+(8\Delta-2K)\lambda^2\nonumber\\
&\hskip4ex+(8\Delta^2-4K\Delta)\lambda+4\Delta^3-3K\Delta^2.\label{eq:43}
\end{align}
(Note that $\rho(\lambda)$ has only three roots. The six eigenvalues of $DF$ are given by the three roots of $DF$, each of multiplicity two. As in the quartic case, this degeneracy corresponds to the rotational invariance of the dynamics.) Inspecting Eq.~(\ref{eq:43}), we find that all eigenvalues have negative real part for sufficiently small $K$ until the a real root crosses $\lambda=0$ at $K_c=4\Delta/3$, which agrees with the results plotted in Fig.~\ref{fig:05} as well as Kuramoto's original self-consistency analysis.

\subsection{Octic Distribution}\label{subsec:0403}

Next we consider the case of the octic distribution, i.e., $g(\omega)=g_n(\omega)$ for $n=4$. From Table~\ref{table:Residue} there are four poles in the lower-half complex plane given by
\begin{align}
\widehat{\omega}_{0/3}&=\pm\frac{\sqrt{2+\sqrt{2}}}{2}\Delta-\frac{\sqrt{2-\sqrt{2}}}{2}\Delta i,\nonumber\\
\widehat{\omega}_{1/2}&=\pm\frac{\sqrt{2-\sqrt{2}}}{2}\Delta-\frac{\sqrt{2+\sqrt{2}}}{2}\Delta i,\label{eq:44}
\end{align}
whose residues are, respectively
\begin{align}
\text{Res}(g_3;\widehat{\omega}_{0/3})=\frac{-(2-\sqrt{2})\mp\sqrt{2}i}{8\pi i},\nonumber\\
\text{Res}(g_3;\widehat{\omega}_{1/2})=\frac{-\sqrt{2}\mp(2-\sqrt{2})}{8\pi i}.\label{eq:45}
\end{align}
Inserting these into Eq.~(\ref{eq:30}) for both poles yields the following system of four complex-valued differential equations:
\begin{widetext}
\begin{align}
\dot{z}_0 &=-\frac{\left(\sqrt{2-\sqrt{2}}-\sqrt{2+\sqrt{2}}i\right)\Delta}{2}z_0+\frac{K}{8}\left[((2-\sqrt{2})-\sqrt{2}i)\left(z_0+z_1+z_2+z_3\right)-\frac{16\left(z_0^*+z_1^*+z_2^*+z_3^*\right)z_0^2}{(2-\sqrt{2})-\sqrt{2}i}\right], \label{eq:46}\\
\dot{z}_1 &=-\frac{\left(\sqrt{2+\sqrt{2}}-\sqrt{2-\sqrt{2}}i\right)\Delta}{2}z_1+\frac{K}{8}\left[(\sqrt{2}-(2-\sqrt{2})i)\left(z_0+z_1+z_2+z_3\right)-\frac{16\left(z_0^*+z_1^*+z_2^*+z_3^*\right)z_1^2}{\sqrt{2}-(2-\sqrt{2})i}\right], \label{eq:47}\\
\dot{z}_2 &=-\frac{\left(\sqrt{2+\sqrt{2}}+\sqrt{2-\sqrt{2}}i\right)\Delta}{2}z_2+\frac{K}{8}\left[(\sqrt{2}+(2-\sqrt{2})i)\left(z_0+z_1+z_2+z_3\right)-\frac{16\left(z_0^*+z_1^*+z_2^*+z_3^*\right)z_2^2}{\sqrt{2}+(2-\sqrt{2})i}\right], \label{eq:48}\\
\dot{z}_3 &=-\frac{\left(\sqrt{2-\sqrt{2}}+\sqrt{2+\sqrt{2}}i\right)\Delta}{2}z_3+\frac{K}{8}\left[((2-\sqrt{2})+\sqrt{2}i)\left(z_0+z_1+z_2+z_3\right)-\frac{16\left(z_0^*+z_1^*+z_2^*+z_3^*\right)z_3^2}{(2-\sqrt{2})+\sqrt{2}i}\right], \label{eq:49}
\end{align}
\end{widetext}
with the Kuramoto order parameter defined by $z=z_0+z_1+z_2+z_3$.

\begin{figure}[htbp]
\centering
\epsfig{file =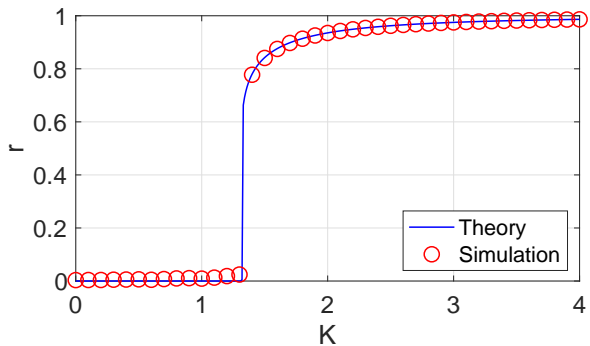, clip =,width=1.0\linewidth }
\caption{{\bf Macroscopic steady-state: Octic distribution}. Steady-state of the macroscopic dynamics measured by the amplitude of the order parameter, $r=|z|$, for the octic distribution, i.e., $n=4$, with $\Delta=1$. Results from a direct simulation of a system of $N=50000$ oscillators is illustrated with red circles, while the analytical prediction obtained by simulating the low dimensional dynamics in Eqs.~(\ref{eq:46}), (\ref{eq:47}), (\ref{eq:48}), and (\ref{eq:49}), where $z=z_0+z_1+z_2+z_3$, is illustrated with a blue curve.}\label{fig:06}
\end{figure}

We again illustrate the utility of capturing the macroscopic dynamics of large systems using the low dimensional dynamics of Eqs.~(\ref{eq:46}), (\ref{eq:47}), (\ref{eq:48}), and (\ref{eq:49}). In Fig.~\ref{fig:06} we plot the degree of synchronization $r=|z|$ vs the coupling strength $K$ obtained first using simulations of the low dimensional system in Eqs.~(\ref{eq:46}), (\ref{eq:47}), (\ref{eq:48}), and (\ref{eq:49}) with $\Delta=1$, denoting results using a blue curve, and second using direct simulations of the system in Eq.~(\ref{eq:01}) with $N=50000$ oscillators whose frequencies are drawn from the octic distribution, i.e., $n=4$, with $\Delta=1$, denoting results with red circles. Again, we note an excellent agreement between the theoretical and directly simulated results.

Next, we illustrate that a linear stability analysis may in this case also reveal the onset of synchronization. Similar to the previous cases, we investigate the stability of the $(z_0,z_1,z_2,z_3)=(0,0,0,0)$ state by calculating the Jacobian matrix of the dynamics after separating into real and imaginary parts, $z_0=x_0+iy_0$, $z_1=x_1+iy_1$, $z_2=x_2+iy_2$, and $z_3=x_3+iy_3$. The resulting $8\times 8$ Jacobian matrix corresponding to $(x_0,y_0,x_1,y_1,x_2,y_2,x_3,y_3)=(0,0,0,0,0,0,0,0)$ is given by
\begin{align}
DF=\frac{\sqrt{2}}{8}\begin{bmatrix}
a & b & c & K & c & K & c & K\\
-b & a & -K & c & -K & c & -K & c\\
K & c & b & a & K & c & K & c\\
-c & K & -a & b & -c & K & -c & K\\
K & -c & K & -c & b & -a & K & -c\\
c & K & c & K & a & b &c & K\\
c & -K & c & -K & c & -K & a & -b\\
K & c & K & c & K & c & b & a
 \end{bmatrix},\label{eq:50}
\end{align}
where $a =(\sqrt{2}-1)K-2\sqrt{2(2-\sqrt{2})}\Delta$, $b=K-2\sqrt{2(2+\sqrt{2})}\Delta$, and $c=(\sqrt{2}-1)K$. The eigenvalues of $DF$ in Eq.~(\ref{eq:50}) are difficult to write down in closed form, and its characteristic polynomial is very complicated, however it can be verified that its roots all have negative real part for sufficiently small $K$ until a double real root passes beomces non-negative at $K_c=\Delta/\sqrt{2-\sqrt{2}}$, which agrees with the results plotted in Fig.~\ref{fig:06} as well as Kuramoto's original self-consistency analysis.

\section{Scaling Properties Near the Onset of Synchronization}\label{subsec:05}

Before concluding we turn our attention to the scaling properties of the order parameter $r=|z|$ near the onset of synchronization. In particular, given the critical coupling strength $K_c$ for the various frequency distributions studied above, how does $r$ increase as $K$ is increased beyond $K_c$. In Kuramoto's original analysis, it was shown that if the frequency distribution $g(\omega)$ is unimodal and symmetric about the mean $\omega=\Omega$ with negative second derivative at the mean, i.e., $g''(\Omega)<0$, then after onset $r$ scales according to a square root of $K-K_c$. This can be observed for the case of Cauchy-distributed frequencies. Recall that for $n=1$ we have $K_c=2\Delta$, and assuming that $0\le K-K_c\ll1$, we have that
\begin{align}
r&=\sqrt{1-\frac{2\Delta}{K}}=\sqrt{\frac{K-K_c}{K}}\nonumber\\
&=\frac{\sqrt{K-K_c}}{\sqrt{K_c}}+\mathcal{O}\left((K-K_c)^{3/2}\right).\label{eq:04:33}
\end{align}
This square root behavior can clearly be seen in Fig.~\ref{fig:01}. However, this analysis breaks down for any distribution that still satisfies the unimodality and symmetry conditions, but for which the second derivative condition fails, i.e., $g''(\Omega)=0$. Note here that for $n\ge2$, the rational distribution $g_n(\omega)$ is an example of a distribution that fails the second derivative condition. In fact, it appears that in Fig.~\ref{fig:04}, Fig.~\ref{fig:05}, and Fig.~\ref{fig:06}, i.e., for $n=2$, $3$, and $4$, that the transition from incoherence to synchronization is sharper than a square-root scaling, and in fact becomes sharper as $n$ increases. We then consider the following questions: Assuming that $0\le K-K_c\ll1$, what then is the scaling properties of $r$? More precisely, given the relationship
\begin{align}
r\appropto(K-K_c)^\gamma,\label{eq:52}
\end{align}
what is the critical scaling exponent $\gamma$ for $0\le K-K_c\ll1$?

\begin{figure}[t]
\centering
\epsfig{file =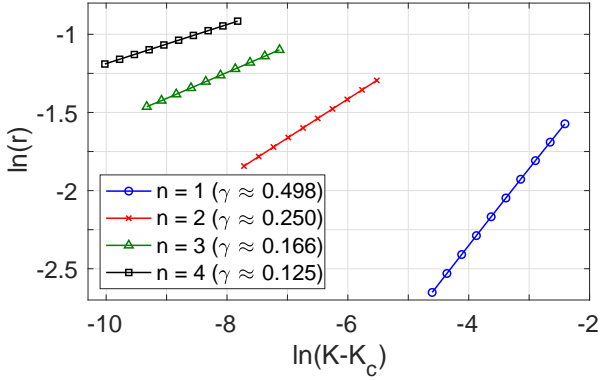, clip =,width=1.0\linewidth }
\caption{{\bf Scaling properties near the onset of synchronization}. For quadratic, quartic, sextic, and octic distributions, the scaling of the order parameter $r=|z|$ near the onset of synchronization, illustrated by plotting $\ln(r)$ vs $\ln(K-K_c)$. Results are obtained by simulating the low dimensional equations for each case and plotted in blue circles, red crosses, green triangles, and black squares, respectively. The slope $\gamma$, representing the critical scaling exponent in $r\propto(K-K_c)^\gamma$ is numerically calculated and given by $\gamma=0.498$, $0.250$, $0.166$, and $0.125$, respectively.}\label{fig:07}
\end{figure}

To shed light on this question we utilize the low dimensional dynamics summarized in Sec.~\ref{sec:04}, which we can simulate numerically. Note that in the absence of a low dimensional description for the macroscopic dynamics, simulations of the full system in Eq.~(\ref{eq:01}) are subject to significant finite-size effects, making such a scaling investigation difficult even for very large systems. Here we proceed by simulating the low dimensional dynamics of the quadratic [Eq.~(\ref{eq:10})], quartic [Eqs.~(\ref{eq:33}) and (\ref{eq:34})], sextic [Eqs.~(\ref{eq:39}), (\ref{eq:40}), and (\ref{eq:41})], and octic [Eqs.~(\ref{eq:46}), (\ref{eq:47}), (\ref{eq:48}), and (\ref{eq:49})] cases over a range of coupling strength larger than, but close to, their respective critical values of $K_c=2\Delta$, $\sqrt{2}\Delta$, $4\Delta/3$, and $\Delta/\sqrt{2-\sqrt{2}}$. (Here we use $\Delta=1$.) In Fig.~\ref{fig:07} we show the results from these simulations, plotting $\ln(r)$ vs $\ln(K-K_c)$ for the quadratic, quartic, sextic, and octic cases in blue circles, red crosses, green triangles, and black squares, respectively. Note that after taking natural logarithms, Eq.~(\ref{eq:52}) becomes
\begin{align}
\ln(r)\approx\gamma\ln(K-K_c)+\text{constant},\label{eq:53}
\end{align}
so that the slopes of the lines in Fig.~\ref{fig:07} reveals the critical exponent $\gamma$. Using a linear least squares best fit to the results using least squares results in critical exponent values approximately given by $\gamma=0.498$, $0.250$, $0.166$, and $0.125$ for the quadratic, quartic, sextic, and octic cases, respectively. 

Regarding the results of these scaling properties, we emphasize that, due to the fact that they are obtained using numerical simulations of the low dimensional dynamics, they represent approximations of the true scaling exponent $\gamma$. However, we make two important comments. First, as the order of the distribution, i.e., $n$, increases, the exponent $\gamma$ decreases, yielding a sharper and sharper transition from incoherence to synchronization. Moreover, recall that in the limit as $n\to\infty$ we recover a uniform natural frequency distribution. It has been previously shown that the uniform case, i.e., the limit $n\to\infty$, results in a first-order transition to synchronization~\cite{Pazo2005PRE}, i.e., a scaling exponent of $\gamma\to0^+$. This interesting property thus fits within the context of what we observe here, connecting the uniform distribution case with our family of rational distributions. 

Second, we note that the numerical approximation for $\gamma$, i.e., $0.498$, $0.250$, $0.166$, and $0.125$ for $n=1$, $2$, $3$, and $4$, respectively, are very close to $1/2$, $1/4$, $1/6$, and $1/8$, respectively. This suggests more than just a decreasing trend in the exponent $\gamma$, but a particular pattern in the precise value of the scaling exponent. Namely, this is strong evidence that suggests that the scaling exponent for the order parameter near onset is given by $\gamma=1/(2n)$, where $2n$ is the order of the rational distribution in Eq.~(\ref{eq:14}). This leads us to wonder whether or not this pattern actually holds generally for the rational distributions considered in this work, i.e., whether the scaling properties truly satisfy $r\appropto\left(K-K_c\right)^{1/(2n)}$.

In fact, we can answer this question in the affirmative by returning to Kuramoto's original self-consistency analysis. Along with the critical onset value $K_c=2/(\pi g(\Omega))$ Kuramoto showed that in the limit $N\to\infty$ the following condition must hold~\cite{Kuramoto}:
\begin{align}
r=rK\int_{-\pi/2}^{\pi/2}\cos^2(\theta)g(Kr\sin(\theta))d\theta.\label{eq:54}
\end{align}
In fact, by evaluating Eq.~(\ref{eq:54}) in the limit $r\to0^+$ the critical onset value can be determined. Moreover, we show here that the critical scaling exponent can be calculate by considering the same limit. Note that for small $r$ the argument of $g$ in Eq.~(\ref{eq:54}) is also small. Using $g_n$ from Eq.~(\ref{eq:14}), we consider the Taylor expansion, which is given by
\begin{align}
g_n(\omega)&=\frac{n\sin(\pi/2n)}{\Delta}\sum_{j=0}^\infty(-1)^j\frac{\omega^{2jn}}{\Delta^{2jn}}\nonumber\\
&=\frac{n\sin(\pi/2n)}{\Delta}\left(1-\frac{\omega^{2n}}{\Delta^{2n}}+\mathcal{O}(\omega^{4n})\right).\label{eq:55}
\end{align}
Inserting this into Eq.~(\ref{eq:54}), eliminating the $r=0$ solution by canceling terms, yields, in the $r\to0^+$ limit
\begin{align}
1&=\frac{n\sin(\pi/2n)K}{\Delta}\left[\int_{-\pi/2}^{\pi/2}\cos^2(\theta)d\theta\right.\nonumber\\
&\hskip6ex-\left.\frac{K^{2n}r^{2n}}{\Delta^{2n}}\int_{-\pi/2}^{\pi/2}\cos^2(\theta)\sin^{2n}(\theta)d\theta\right].\label{eq:56}
\end{align}
Rearranging Eq.~(\ref{eq:56}) to solve for $r$ and using that $K_c=2/\pi g(0)=2\Delta/\pi n\sin(\pi/2n)$, we find that
\begin{align}
r=\frac{\Delta}{K(2AK)^{2n}}(K-K_c)^{1/(2n)},\label{eq:57}
\end{align}
where $A=\pi^{-1}\int_{-\pi/2}^{\pi/2}\cos^2(\theta)\sin^{2n}(\theta)d\theta$. 

This shows that, for the natural frequency distribution $g_n(\omega)$ given in Eq.~(\ref{eq:14}), near the onset of synchronization the order parameter truly does scale like $r\appropto(K-K_c)^{1/(2n)}$. Moreover, this particular pattern unifies the scaling behavior for the Kuramoto model with rational frequency distributions with the case of the uniform distribution. In particular, when natural frequencies are uniformly distributed the onset of synchronization occurs via a discontinuous jump, i.e., a first-order phase transition~\cite{Pazo2005PRE}. In terms of a scaling exponent this corresponds to a scaling exponent $\gamma\to0^+$. On the other hand, we have shown that in the limit $n\to\infty$ the family of rational distributions we consider here recovers the uniform distribution, in which case the scaling exponent is given by $\gamma=\lim_{n\to\infty}1/(2n)=0$, which is in agreement with these previous results. In particular, as we choose a larger and larger value of $n$ for the rational distribution $g_n(\omega)$, we obtain a steeper and steeper transition to synchronization, getting closer and closer to a truly first-order phase transition.

\section{Discussion}\label{sec:06}

In this paper we have investigated the low dimensional dynamics of the Kuramoto model in the context of different natural frequency distributions. By using the dimensionality reduction of Ott and Antonsen the dynamics of an infinitely large system of coupled oscillators can be reduced to a low dimensional system describing the dynamics of the Kuramoto order parameter. However, in order to close the low dimensional dynamics it is assumed in the vast majority of works that use the so-called Ott and Antonsen anstaz that the natural frequencies are drawn from a Cauchy distribution. Here we have generalized this setup, investigating the effects of using frequency distributions from a broader family of rational distributions. 

Denoting the order of the distribution as the order of the polynomial characterizing the distribution (i.e., the order of the polynomial in the denominator of the distribution), we close the dynamics obtained by the dimensionality reduction of Ott and Antonsen by integration in the complex plane using the poles and respective residues of the frequency distributions. In particular, the distribution of order $2n$ has precisely $2n$ simple poles -- $n$ in the upper-half complex plane and $n$ in the lower-half complex plane -- which allows us to close the low dimensional dynamics in a system of $n$ complex-valued differential equations defining the evolution of $n$ pseudo order parameters from which we easily calculate the Kuramoto order parameter. After outlining the general framework, we consider the specific cases of the quartic, sextic, and octic distributions, i.e., $n=2$, $3$, and $4$, (where the Cauchy distribution corresponds to the quadratic case of $n=1$). For each case we demonstrate that the low dimensional equations accurately describe the macroscopic dynamics of the original system, and moreover the low dimensional equations can be used to calculate the onset of synchronization via a bifurcation analysis. Finally, we used the low dimensional equations to perform an analysis of the scaling properties of the order parameter near the onset of synchronization, calculating the critical scaling exponent for different choices of frequency distributions. We then corroborated these results with additional calculations utilizing Kuramoto's original self-consistency framework, showing that the critical scaling exponent of the order parameter near onset is precisely the inverse of the order of the rational distribution used. In addition to identifying this particular pattern, this also puts into context the case of uniformly-distributed frequencies, for which case the transition is discontinuous, and corresponds to the limit of a rational distribution with infinite order.

Before closing, we note that beyond the analysis of the critical coupling strength for the onset of synchronization and the scaling properties of the order parameter near onset, the low dimensional analysis provided above allows for simple. method for calculating the degree of synchronization of the system as measured by the order parameter. Note in particular, that Kuramoto's self-consistency analysis gives the order parameter implicitly in a complex integral equation that, in general, cannot be evaluated analytically. Thus, a complicated quadrature iterations is required to compute the order parameter using this method. In contrast, using the low dimensional equations presented here one may simply solve a relatively simple system of differential equations to obtain an accurate value. (Moreover, this can be done with a wide variety of numerical methods for solving differential equations.)

Lastly, we note that the work presented here is strictly for the continuum limit of infinitely-large oscillator systems. It remains to be seen how the choice of frequency distributions as considered here would affect finite-size effects in the macroscopic system dynamics, in particular finite-size fluctuations and finite-size scaling of the order parameter~\cite{Buice2007PRE,Coletta2017PRE}. Such investigations may be fruitbful avenues of future research.

\appendix
\section{Poles and Residues of $g_n(\omega)$}\label{app:A}

Here we present the full set of residues $\text{Res}(g_n;\widehat{\omega})$ for the quadratic, quartic, sextic, and octic cases, i.e., $n=1$, $2$, $3$, and $4$, in Table~\ref{table:Residue}. In the first column we denote $n$ and the distribution order, followed by the location of the poles of $g_n(\omega)$, $\widehat{\omega}$, in the second column, and finally the poles' respective residues in the third column. For convenient use with the Cauchy residue theorem in the following section, we report residues in the form of a complex number divided by a multiple of $2\pi i$. We will use these results in the following section, but for now we may observe a variety of patterns in the residues, e.g., for complex conjugate poles the numerators of the residues all have the same imaginary part, but opposite real parts.

\begin{widetext}

\begin{table}[htpb]
\centering
 \caption{Summary of the poles of $g_n(\omega)$, $\widehat{\omega}$, and their respective residues, $\text{Res}(g_n;\widehat{\omega})$ for $n=1$, $2$, $3$, and $4$, i.e., the quadratic, quartic, sextic, and octic distributions.}
 \label{table:Residue}
\begin{tabular}{|c|c|c|}
    \hline
    $n$ (Distribution order) & poles of $g_n(\omega)$, $\widehat{\omega}$ & Residue, $\text{Res}(g_n;\widehat{\omega})$  \\ 
    \hline
    \hline
    1 (quadratic) & $\Delta i$ & $1/(2\pi i)$ \\ 
     & $-\Delta i$ & $-1/(2\pi i)$ \\ \hline
    2 (quartic) & $\frac{1}{\sqrt{2}}\Delta+\frac{1}{\sqrt{2}}\Delta i$ & $(1-i)/(4\pi i)$ \\
     & $-\frac{1}{\sqrt{2}}\Delta+\frac{1}{\sqrt{2}}\Delta i$ & $(1+i)/(4\pi i)$ \\
     & $-\frac{1}{\sqrt{2}}\Delta-\frac{1}{\sqrt{2}}\Delta i$ & $(-1+i)/(4\pi i)$ \\
     & $\frac{1}{\sqrt{2}}\Delta-\frac{1}{\sqrt{2}}\Delta i$ & $(-1-i)/(4\pi i)$ \\  \hline
     3 (sextic) & $\frac{\sqrt{3}}{2}\Delta+\frac{1}{2}\Delta i$ & $(1-\sqrt{3}i)/(8\pi i)$ \\
     & $\Delta i$ & $1/(4\pi i)$ \\
     & $-\frac{\sqrt{3}}{2}\Delta+\frac{1}{2}\Delta i$ & $(1+\sqrt{3}i)/(8\pi i)$ \\
     & $-\frac{\sqrt{3}}{2}\Delta-\frac{1}{2}\Delta i$ & $(-1+\sqrt{3}i)/(8\pi i)$ \\
     & $-\Delta i$ & $-1/(4\pi i)$ \\
     & $\frac{\sqrt{3}}{2}\Delta-\frac{1}{2}\Delta i$ & $(-1-\sqrt{3}i)/(8\pi i)$ \\
     \hline
     4 (octic) & $\frac{\sqrt{2+\sqrt{2}}}{2}\Delta+\frac{\sqrt{2-\sqrt{2}}}{2}\Delta i$ & $\left[\left(2-\sqrt{2}\right)-\sqrt{2}i\right]/(8\pi i)$ \\
     & $\frac{\sqrt{2-\sqrt{2}}}{2}\Delta+\frac{\sqrt{2+\sqrt{2}}}{2}\Delta i$ & $\left[\sqrt{2} - \left(2-\sqrt{2}\right)i\right]/(8\pi i)$ \\
     & $-\frac{\sqrt{2-\sqrt{2}}}{2}\Delta+\frac{\sqrt{2+\sqrt{2}}}{2}\Delta i$ & $\left[\sqrt{2} + \left(2-\sqrt{2}\right)i\right]/(8\pi i)$ \\
     & $-\frac{\sqrt{2+\sqrt{2}}}{2}\Delta+\frac{\sqrt{2-\sqrt{2}}}{2}\Delta i$ & $\left[\left(2-\sqrt{2}\right)+\sqrt{2}i\right]/(8\pi i)$ \\
     & $-\frac{\sqrt{2+\sqrt{2}}}{2}\Delta-\frac{\sqrt{2-\sqrt{2}}}{2}\Delta i$ & $\left[-\left(2-\sqrt{2}\right)+\sqrt{2}i\right]/(8\pi i)$ \\
     & $-\frac{\sqrt{2-\sqrt{2}}}{2}\Delta-\frac{\sqrt{2+\sqrt{2}}}{2}\Delta i$ & $\left[-\sqrt{2}+\left(2-\sqrt{2}\right)i\right]/(8\pi i)$ \\
     & $\frac{\sqrt{2-\sqrt{2}}}{2}\Delta-\frac{\sqrt{2+\sqrt{2}}}{2}\Delta i$ & $\left[-\sqrt{2}-\left(2-\sqrt{2}\right)i\right]/(8\pi i)$ \\
     & $\frac{\sqrt{2+\sqrt{2}}}{2}\Delta-\frac{\sqrt{2-\sqrt{2}}}{2}\Delta i$ & $\left[-\left(2-\sqrt{2}\right)-\sqrt{2}i\right]/(8\pi i)$ \\
     \hline
\end{tabular}
\end{table}

\end{widetext}

%\acknowledgements
%This work was supported by the James S. McDonnell Foundation (PSS) and ARO Grant W911NF1210101 (EO).

\bibliographystyle{plain}

\end{document}